\documentclass[twocolumn,amsmath,amssymb]{snp}
\usepackage{graphicx}
\usepackage{dcolumn}
\usepackage{bm}
\begin{document}

\title{\Large Effect of nuclear compressibility on the fragmentation \\ in peripheral Au+Au collisions at 35 AMeV}

\author{\large Yogesh K. Vermani}
\affiliation{Dept. of Physics, ITM University, Gurgaon-122017,
INDIA} \email{yugs80@gmail.com}

\author{\large Rajiv Chugh}
\affiliation{Dept. of Physics, Panjab University,
Chandigarh-160014, INDIA}
\author{Aman D. Sood}
\affiliation{SUBATECH - Ecole des Mines de Nantes 4, F-44072
Nantes, Cedex 03, FRANCE}


\maketitle

\noindent {\large{\bf Introduction}}

The heavy-ion collisions in intermediate energy regime can provide
important clues about the nature of baryonic matter equation of
state (EoS) with nucleon density varying between $1/3$ and 3 times
the saturation density $\varrho_{o}(=0.17~fm^{-3})$ of cold nuclei
\cite{stok}. At higher bombarding energies, momentum dependence of
\emph{n-n} interaction as well as in-medium scattering
cross-section affect considerably the phenomenon of collective
flow and multifragmentation \cite{verm, mages}. In Fermi energy
domain, fragmentation of hot nucleus has been given attention in
recent times \cite{rkp}, where \emph{quantum molecular dynamics}
(QMD) model coupled with advanced clustering subroutine namely
\emph{simulated annealing clusterization algorithm} (\emph{SACA})
is reported to explain the $^{16}O+^{80}Br$ fragmentation data at
incident energies $T_{lab}$=50-200 AMeV. For the present work, we
aim to address the problem of nuclear EoS by studying the
fragmentation in peripheral
Au+Au collisions at $T_{lab}=35$ AMeV employing a soft EoS ($\kappa=200 MeV$) and a hard Eos ($\kappa=380 MeV$). \\

\noindent {\large{\bf The Model}}

To generate the phase space of nucleons and see the effect of
different nuclear compressibilities, we employ quantum molecular
dynamics (QMD) model \cite{hart}. In this model, phase space of
nucleons is calculated via Hamilton's equations of motion:
\begin{eqnarray}
\dot{{\bf p}_i}=& -\{{\bf r}_i, {\cal H}\}, \\
\dot{{\bf r}_i}=&\{{\bf p}_i, {\cal H}\}.
\end{eqnarray}
Here ${\cal H}$ is the total Hamiltonian of the system of
$A_{P}+A_{T}$ nucleons. These equations are solved after fixed
time interval $ \bigtriangleup t$ chosen to be very small.

The phase space of nucleons is clusterized using improvised
version of \emph{SACA} \cite{esaca} labeled as \emph{SACA (2.1)}
where binding energy of each cluster $\zeta_{f}$ is checked to be
greater than $E_{bind}$ given as
\begin{eqnarray}
& E_{bind}= a_{v}A_f - a_{s}A_{f}^{2/3} - a_{c}
\frac{Z_{f}(A_{f}-1)}{A_{f}^{1/3}} - \nonumber \\
& a_{sym}\frac{(A_f -2Z_{f})^{2}}{A_{f}} (\pm,0)~a_{p}\frac
{(1-e^{-A_{f}/30})} {A_{f}^{1/2}},
\end{eqnarray}
with $A_{f}$ and $Z_{f}$ as mass and charge of a cluster. The last
term (\emph{i.e.} pairing energy term) is taken to be +ve for
even-even nuclei, -ve for odd-odd nuclei and zero for odd $A_{f}$
nuclei. \\

\noindent {\large{\bf Results and Discussion}}

We display in Fig. 1, the calculations with soft and hard
equations of state for Au(35 AMeV)+Au collisions at reduced impact
parameters $\hat{b}=0.55$ and $0.85$, respectively. One obtains a
clear difference in the fragment observables due  to choice of
nuclear compressibility. Asymptotic size of $A^{max}$ and
multiplicity of heavier clusters [$3\leq Z \leq80$] saturate as
early as 100 fm/c when system has just evolved from violent phase.
With soft EoS, a larger number of heavier fragments are produced,
thereby, decreasing the size of $A^{max}$, while hard \emph{n-n}
collisions result into emission of free nucleons mostly. Earlier
recognition of cluster configuration suggest that \emph{SACA
(2.1)} is well suited for the study of heavy ion reactions in low
energy regime.
\begin{figure} [!t]
\centering \setlength{\abovecaptionskip}{-1.0cm}
\setlength{\belowcaptionskip}{0.5cm} \vskip -0.50cm
\includegraphics [width=62mm]{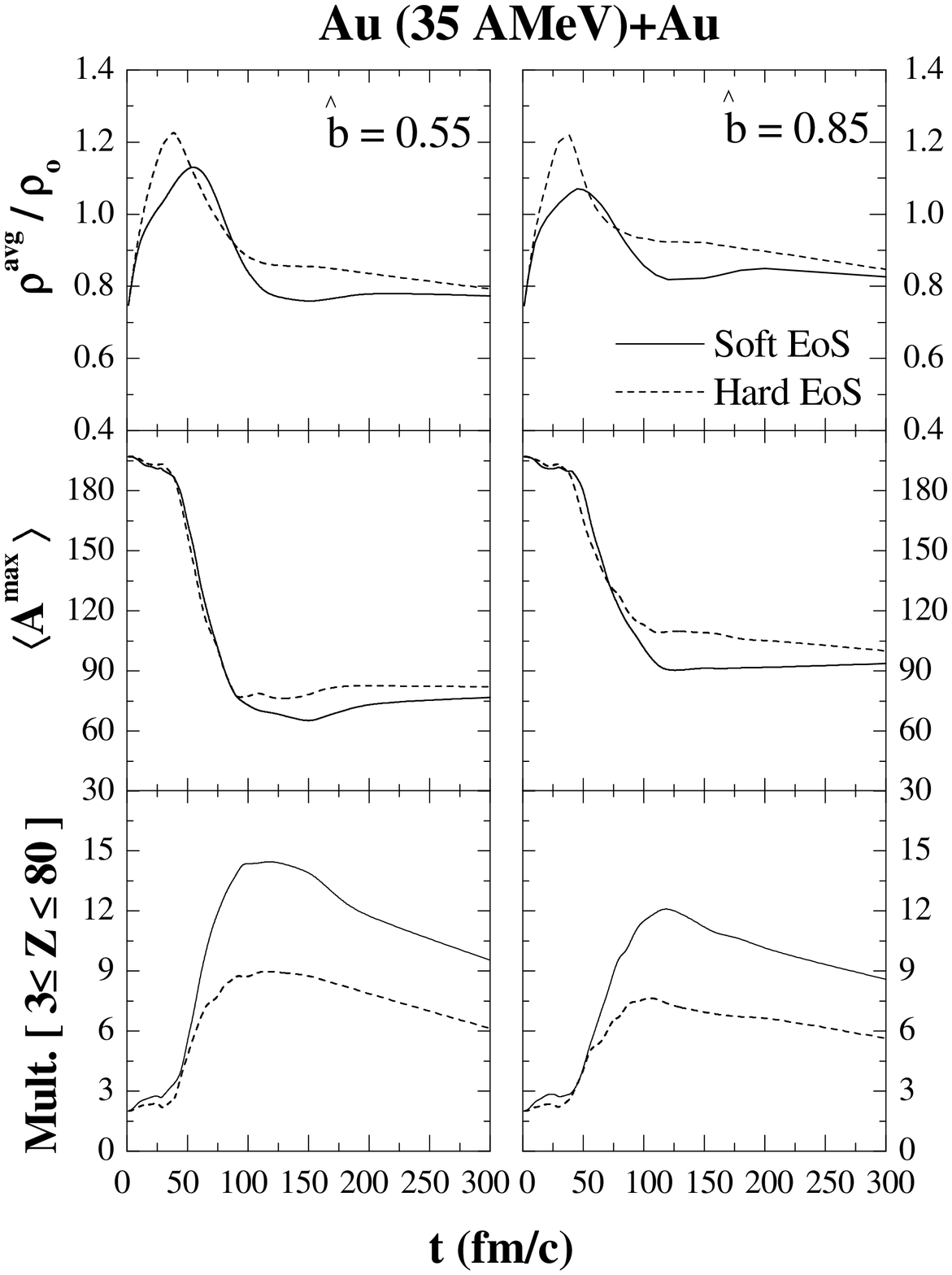}
\vskip 0.5cm \caption {The time evolution of mean nucleon density
$\rho^{avg}$ (top), size of heaviest fragment $A^{max}$ (middle),
and multiplicity of fragments with charge $3\leq Z \leq80$
(bottom).}

\vskip -0.95cm
\includegraphics [scale=0.30]{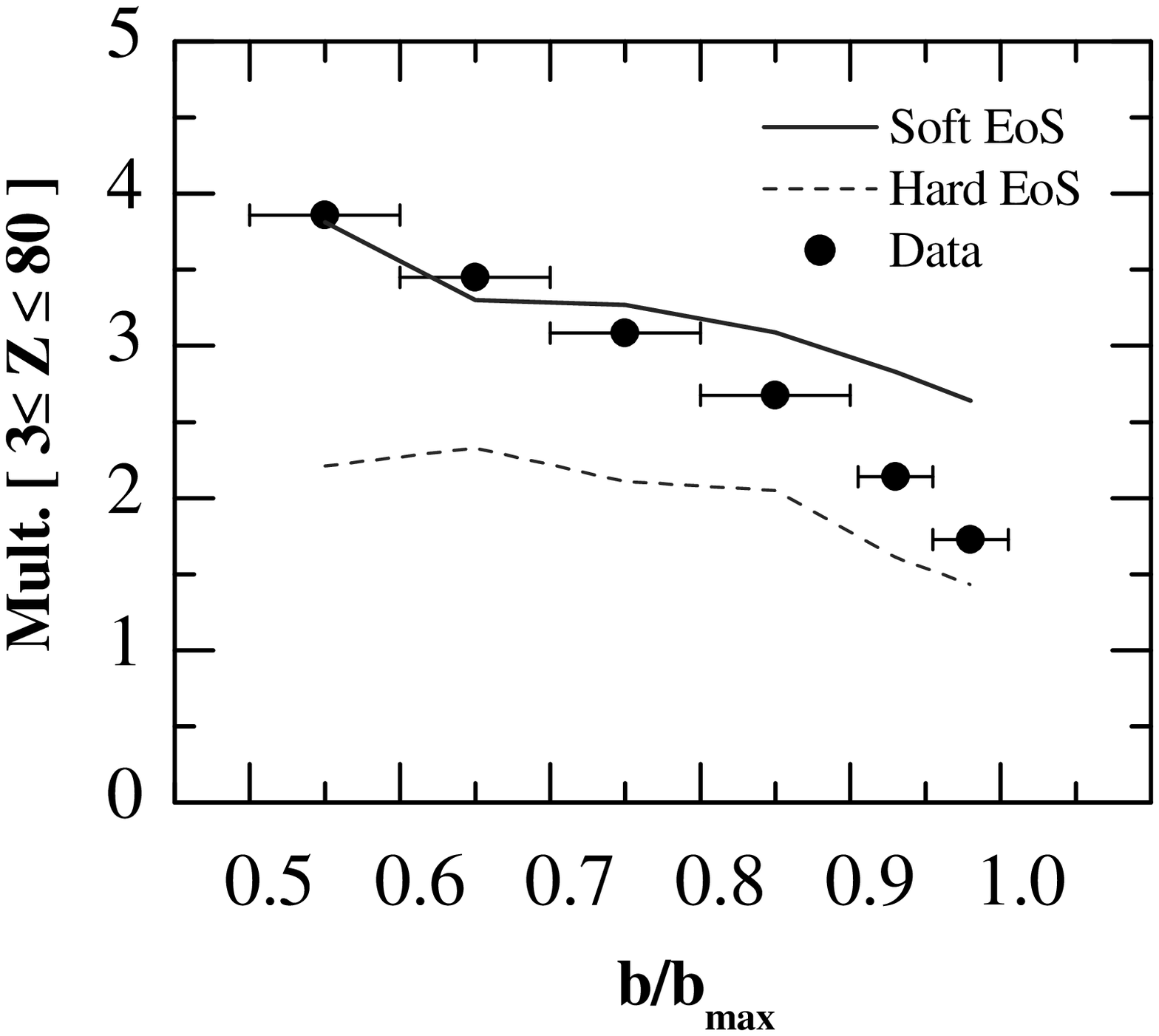}
\vskip -1.2cm \caption {The impact parameter dependence of
multiplicity of fragments with charge $3\leq Z \leq80$ obtained
with a `soft' EoS (solid line) and a `hard' EoS (dashed line) in
Au(35 AMeV)+Au collision. Filled circles depict the experimental
data points \cite{ago} (Preliminary Results).}
\end{figure}
Next, we turn to comparison of model calculations for the
multiplicity of fragments [$3\leq Z \leq80$] obtained from decay
of spectator matter moving in forward direction with rapidity
$y>0.5~y_{beam}$ in the c.m. frame with experimental data taken
with Miniball-Multics array \cite{ago}. Figure 2 depicts the
results of QMD simulations for soft and hard interactions at 100
fm/c as a function of reduced impact parameter $b/b_{max}$. The
mean multiplicity of fragments falls with increase in the impact
parameter. This can be understood in terms of lesser transfer of
energy from hot participant zone to spectator region at peripheral
geometries. One can clearly see that soft EoS accurately
reproduces the experimental trend of fragment multiplicity as a
function of impact parameter. The hard EoS on the contrary, seems
too explosive to explain the data.

This study shows that \emph{SACA (2.1)} can well describe the
early dynamics of reactions at low energy. The predictions with
QMD model for the fragment yield in peripheral Au+Au collisions
favor \emph{soft} nature of baryonic matter \cite{yugs}. \\


\end{document}